\pdfoutput=1

\documentclass[epj]{svjour}

\usepackage[utf8]{inputenc}
\usepackage{amsmath, amssymb}
\usepackage{textcomp}
\usepackage{graphicx}
\usepackage{ellipsis}
\usepackage{cite}
\usepackage{authblk}
\usepackage{microtype}
\usepackage{color}

\title{Final-State Interactions in the Process \boldmath{$\pol pp\to
  pK^{\!+}\!\Lambda$}}
\titlerunning{Final-State Interactions in the Process $\pol pp\to
  pK^{\!+}\!\Lambda$}
\subtitle{The COSY-TOF Collaboration}

\author{M.~Röder\thanks{m.roeder@fz-juelich.de}\inst{1,2} \and
  E.~Borodina\inst{1,2}\and
  H.~Clement\inst{6,7} \and
  E.~Doroshkevich\inst{6,7} \and
  R.~Dzhygadlo\inst{1,2} \and
  K.~Ehrhardt\inst{6,7} \and
  A.~Erhardt\inst{6,7} \and
  W.~Eyrich\inst{5} \and
  W.~Gast\inst{1,2} \and
  A.~Gillitzer\inst{1,2} \and
  D.~Grzonka\inst{1,2} \and
  J.~Haidenbauer\inst{1,2,3} \and
  C.~Hanhart\inst{1,2,3} \and
  F.~Hauenstein\inst{1,2,5} \and
  P.~Klaja\inst{1,2,5} \and
  L.~Kober\inst{5} \and
  K.~Kilian\inst{1,2} \and
  M.~Krapp\inst{5} \and
  M.~Mertens\inst{1,2}\and
  J.~Ritman\inst{1,2} \and
  E.~Roderburg\inst{1,2} \and
  W.~Schroeder\inst{2} \and
  T.~Sefzick\inst{1,2} \and
  A.~Sibirtsev\inst{8} \and
  P.~Wintz\inst{1,2} \and
  P.~Wüstner\inst{2,4}}

\institute{
  Institut für Kernphysik, Forschungszentrum Jülich, 52428 Jülich, Germany \and
  Jülich Center for Hadron Physics, Forschungszentrum Jülich, 52428 Jülich, Germany  \and
  Institute for Advanced Simulation, Forschungszentrum Jülich, 52428 Jülich, Germany \and
  Zentralinstitut für Elektronik, Forschungszentrum Jülich, 52428 Jülich, Germany  \and
  Friedrich-Alexander-Universität Erlangen-Nürnberg, 91058 Erlangen, Germany \and
  Physikalisches Institut der Universität Tübingen, Auf der Morgenstelle 14, 72076 Tübingen, Germany  \and
  Kepler Center for Astro and Particle Physics, University of Tübingen, Auf der Morgenstelle 14, 72076 Tübingen, Germany  \and
  Helmholtz-Institut für Strahlen- und Kernphysik, Nussallee 14-16 53115 Bonn, Germany}

\authorrunning{Matthias Röder et al.}

\newcommand{\pol}[1]{\mathaccent"017E{#1}}
\newcommand{\degree}{\ensuremath{^{\circ}}}
\newcommand{\eV}{\ensuremath{\text{e\kern -0.09em V}}}
\newcommand{\MeV}{\ensuremath{\text{M}\eV}}
\newcommand{\GeV}{\ensuremath{\text{G}\eV}}
\newcommand{\MeVc}{\ensuremath{\MeV\kern -0.15 em/c}}
\newcommand{\MeVcc}{\ensuremath{\MeV\kern -0.15 em/c^{2}}}
\newcommand{\GeVc}{\ensuremath{\GeV\kern -0.15 em/c}}
\newcommand{\GeVcc}{\ensuremath{\GeV\kern -0.15 em/c^{2}}}
\newcommand{\Kaonp}{\ensuremath{K\kern -0.09 em ^{+}\kern -0.07 em }}
\newcommand{\chisquare}{\ensuremath{\chi^{2}}}
\newcommand{\OLDKaonp}{\ensuremath{K\kern -0.09 em ^{+}}}
\newcommand{\kl}{\ensuremath{K\kern -0.1 em \Lambda}}
\newcommand{\pkl}{\ensuremath{p\OLDKaonp\kern -0.2 em \Lambda}}
\newcommand{\pks}{\ensuremath{p\OLDKaonp\kern -0.1em \Sigma^{0}}}
\newcommand{\pptopkl}{\ensuremath{\pol{p}p \to\pkl}}
\newcommand{\pptopkstopklgamma}{\ensuremath{\pol{p}p \to \pks \to
    \pkl\gamma}}
\newcommand{\pptopkltopkppi}{\ensuremath{\pol{p}p \to \pkl
    \to p\OLDKaonp \kern -1pt  p\pi^{-}}}
\newcommand{\pkltopkppi}{\ensuremath{\pkl\to p\OLDKaonp \kern -1pt  p\pi}}

\newcommand{\nstar}{\ensuremath{N^{*}}\!}
\newcommand{\mum}{\textmu m}

\definecolor{Emph}{rgb}{1,0,0}  

\begin{document}

\abstract{ The possibility to determine the $p\Lambda$ scattering
  length from the final-state interaction in the reaction
  \mbox{$\pptopkl$} is investigated experimentally. From a
  transversely polarized measurement, the \Kaonp{} analyzing power
  ($A_{N}$) which, in principle, allows one to extract the spin
  triplet scattering length is studied. An unexpected energy
  dependence of the forward/backward symmetric part of $A_{N}$ is
  found. The influence of \nstar{} resonances on the $p\Lambda$
  invariant mass spectrum is investigated by exploiting the large
  acceptance for the process \mbox{$\pptopkltopkppi$} and is found to
  be the main source of uncertainty for determining the $p\Lambda$
  scattering length. }


\maketitle

\section{Introduction}
\label{sec:introduction}

While nucleon-nucleon scattering has been precisely measured and
accurately described up to beam kinetic energies of
$3\,\GeV$\cite{Arndt:2000xc}, the situation is much worse for
hyperon-nucleon scattering. Recently, a dispersion relation technique
was developed by Gasparyan et al.\cite{Gasparyan:2003cc} that allows
one to extract the $p\Lambda$ scattering length from final-state
interactions in processes with high momentum transfer like
\mbox{\pptopkl{}}. It requires data where the $p\Lambda$ system is in
a specific total spin (S=0 or S=1). In an experiment with transversely
polarized beam, the symmetric component of the \Kaonp{} analyzing
power ($A_{N}$) can be used to disentangle the spin-triplet component
of the $p\Lambda$ interaction. Then, the scattering length can be
extracted from the shape of the corresponding $p\Lambda$ invariant
mass spectrum.

\nstar{} resonances were found to have a significant influence on the
$pK\Lambda$ production cross sections\cite{AbdElSamad:2010tz,
AbdelBary:2010pc}, see also \cite{Sibirtsev:2005mv}. In
Ref.~\cite{Gasparyan:2003cc} it is pointed out that a Dalitz plot
analysis should be performed to check whether the area of $p\Lambda$
final-state interaction is overlapping with those resonance structures
in order to assess the applicability of the extraction method.

Due to its unique feature of 4$\pi$ acceptance for the final state,
the COSY-TOF spectrometer is ideally suited to measure the
$\pptopkl{}$ reaction and provide the data needed for the extraction
of the scattering length and the Dalitz plot analysis. It has recently
been upgraded for improved event reconstruction capabilities. These
should allow to achieve an experimental precision that competes with
the accuracy of the extraction method. Additionally, the polarized
proton beam from the COSY accelerator gives access to the polarization
observable $A_{N}$.

In this paper we apply the extraction method to data taken at a beam
momentum of $2.95\,\GeVc$, corresponding to an excess energy of
203.7\,MeV. Systematic effects due to the influence of
\nstar{} resonances are quantified by separately analyzing different
regions of the Dalitz plot. We observe an unexpected energy dependence
of the $K^{+}$ analyzing power, and consequently it is not possible to
determine the spin-triplet scattering length at this beam momentum,
with the amount of data available. Nevertheless, this finding is
interesting and further experimental and theoretical studies are
necessary.

\section{Method to Determine the Spin-Triplet Scattering Length (\boldmath{$a_{t}$})}
\label{sec:spintripl}

The $p\Lambda$ interaction in the final state leads to an enhancement
of the production cross section for $p\Lambda$ invariant masses
($m_{p\Lambda}$) near threshold.  In Ref.~\cite{Gasparyan:2003cc} a
dispersion relation is derived that connects the $p\Lambda$ scattering
length to an integral over $m_{p\Lambda}$. The spectrum is integrated
from $m_{0}=m_{p}+m_{\Lambda}$ up to
$m_{\text{max}}=m_{0}+40$\,\MeVcc{}, because the $p\Lambda$ system is
required to be in an S-wave. From simulations it is argued that, with
this limit, the accuracy of the method is $0.3\,\text{fm}$.

The application of the method requires observables where the
$p\Lambda$ system is in a specific total spin state. In Appendix B of
Ref.~\cite{Gasparyan:2003cc} it is pointed out that from a measurement
with a transversely polarized beam observables are accessible to
which only the spin-triplet part of the production amplitude
contributes. Therefore, the spin-triplet scattering length $a_{t}$ can
be determined. To show that, the \Kaonp{} analyzing power is expanded
in terms of the associated Legendre Polynomials $P^{m}_{\ell}$ of
degree $\ell$ and order $m$ \cite{Hanhart:2003pg}:
\begin{equation}
  \label{eq:101}
   \begin{split}
   &A_{N}(\cos\theta^{*}_{\!K}\!, m_{p\Lambda})
   \frac{d^{2}\sigma}{d\cos\theta^{*}_{\!K}dm_{p\Lambda}} 
  \\&=  
   \alpha(m_{p\Lambda}) P_{1}^{1}(\cos\theta^{*}_{\!K})
   + \beta(m_{p\Lambda}) P_{2}^{1}(\cos\theta^{*}_{\!K}) + \dots{}\ .
  \end{split}
\end{equation}
Here, $d^{2}\sigma/d\cos\theta^{*}_{\!K}dm_{p\Lambda}$ is the double
differential production cross section and $\theta^{*}_{\!K}$ is the
polar angle of the \Kaonp{} in the center-of-mass system. The function
$P_{1}^{1}$ ($ P_{2}^{1}$) is forward-backward symmetric
(antisymmetric). Higher order contributions to Eq.~(\ref{eq:101}) turn
out to be negligible in the analysis of the data from the current
experiment. This implies that the coefficient $\alpha$ results from
the interference of amplitudes that correspond to the $\Kaonp{}$ being
in an s- and p-wave, respectively, while the coefficient $\beta$ is
due to an interference of the s- and d-wave amplitudes
\cite{Hanhart:2003pg}.

Under the assumption that the $p\Lambda$ system is in an S-wave, only
spin-triplet amplitudes contribute to $\alpha$ (see Appendix B
of \cite{Gasparyan:2003cc} for details). This follows from the
fact that different spin states in the final state do not interfere 
and that, in addition, for a spin zero $p\Lambda$ final state, only
even \Kaonp{} partial waves are possible\footnote{For a $p\Lambda$
  system in an S-wave and in a spin zero state one has for the total
  angular momentum $J_{\rm tot}=\ell_K$, with $\ell_K$ for the angular
  momentum of the \Kaonp{} with respect to the $p\Lambda$ system. The
  parity of the final state thus reads $\pi_{f}=(-1)^{\ell_{K}+1}$.
  Because the parity of the initial state is given by the relative
  angular momentum $L_{i}$ of the protons as $\pi_{f}=(-1)^{L_{i}}$,
  it follows from parity conservation that $L_{i}\neq J_{\text{tot}}$.
  Therefore, the initial state can only be in spin 1 state. Because
  then the Pauli principle demands that $(-1)^{L_{i}} = -1$, it
  follows from $L_{i}\neq\ell_{K}$ that $\ell_{K}$ has to be even. }.
Consequently, there is no contribution of the $p\Lambda$ singlet state 
to the \Kaonp{} p-wave, hence to $\alpha$. Therefore, the integral
\begin{equation}
  \begin{split}
    &\int_{-1}^{+1}\!d\cos\theta^{*}_{\!K}
    \frac{A_{N}(\cos\theta^{*}_{\!K}\!, m_{p\Lambda})}{p_{p\Lambda}}
    \frac{d^{2}\sigma}{d\cos\theta^{*}_{\!K}dm_{p\Lambda}} \\ & =
    \alpha (m_{p\Lambda}) \frac{\pi}{2}  \ ,
  \end{split}
  \label{eq:102}
\end{equation}
contains only contributions from $p\Lambda$ spin-triplet states
because the forward-backward antisymmetric term proportional to
$\beta$ cancels out.  Here, $p_{p\Lambda}$ denotes the relative
momentum between $p$ and $\Lambda$. 

Using the parametrization
\begin{equation}
  \label{eq:103}
 |\alpha(m_{p\Lambda})| = \exp\left[ C_{0} + \frac{C_{1}^{2}}{(m_{p\Lambda}^{2}-C_{2}^{2})}  \right],
\end{equation}
with the free parameters $C_i$, the spin-triplet scattering length
$a_{t}$ can be obtained from:
\begin{equation}
  \label{eq:46}
  \begin{split}
    &a_{t}(C_{1}, C_{2}) = -\frac{\hbar c}{2}C_{1}^{2}\\
    &\times \sqrt{\left(\frac{m_{0}^{2}}{m_{p}m_{\Lambda}}\right)
      \frac{(m_{\text{max}}^{2} -m_{0}^{2}) }{ (m_{\text{max}}^{2}-
        C_{2}^{2}) (m_{0}^{2}-C_{2}^{2})^{3} }}.
  \end{split}
\end{equation}
The independence of the scattering length from $C_{0}$ reflects the
fact that only the \emph{shape} induced by the final-state interaction 
plays a role --- this is why already the proportionality of the spin-triplet 
scattering amplitude to $|\alpha(m_{p\Lambda})|$ alone is sufficient.

Direct measurements of $d\sigma/d\cos\theta^{*}_{\!K}$ have demonstrated that 
there is practically no dependence of this quantity on $\theta^{*}_{\!K}$ 
\cite{AbdelBary:2010pc,mythesis}. Thus, we can simplify the formalism in
the application to the present experiment. Specifically, instead of 
Eq.~(\ref{eq:101}) we can use 
\begin{equation}
  \label{eq:401}
   \begin{split}
   &A_{N}(\cos\theta^{*}_{\!K}\!, m_{p\Lambda})
  \\&=  
  \bar\alpha(m_{p\Lambda}) P_{1}^{1}(\cos\theta^{*}_{\!K}) + \bar\beta(m_{p\Lambda}) P_{2}^{1}(\cos\theta^{*}_{\!K})
  \ .
  \end{split}
\end{equation}
The quantities $\alpha$ and $\bar \alpha$ are then simply related by 
\begin{equation}
\alpha(m_{p\Lambda}) = \bar \alpha(m_{p\Lambda}) \cdot |M(m_{p\Lambda})|^{2} ,
\end{equation}
where $|M(m_{p\Lambda})|^{2}$ is proportional to the angular- and spin-averaged
enhancement of the production cross section, as discussed above:
\begin{equation}
|M(m_{p\Lambda})|^{2} \propto \frac{1}{p_{p\Lambda}} \frac{d\sigma}{dm_{p\Lambda}} .
\label{eq:7}
\end{equation}
As a consequence we can separately determine the quantities
$\bar \alpha(m_{p\Lambda})$ and $|M(m_{p\Lambda})|^{2}$ and then use
their product in Eqs.~(\ref{eq:103}-\ref{eq:46}).

\section{Experimental Method}
\label{sec:method}

\subsection{Experimental Setup}
\label{sec:experimental-setup}

\begin{figure}
  \centering
  \includegraphics[width=\linewidth]{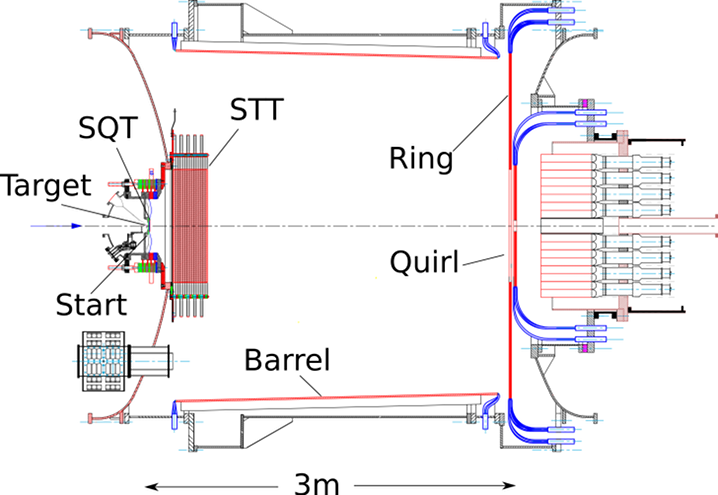}
  \caption{Schematic view of the COSY-TOF detector, including the new
    Straw Tube Tracker (STT) and Silicon Quirl Telescope (SQT).}
  \label{fig:tof}
\end{figure}

The new experimental setup of COSY-TOF is sketched in
Fig.~\ref{fig:tof}. Segmented scintillators close to the target and at
the inner side of the main vacuum vessel are used for triggering, time
of flight and $dE/dx$ measurements. New detection subsystems have been
installed to improve the event reconstruction precision and
efficiency: the Silicon Quirl (SQT) close to the target and the Straw
Tube Tracker (STT) \cite{bib:stt:des:wintzaip} inside the main vacuum
vessel. The STT has nearly $4\pi$ acceptance for the $pp\to
pK^{+}\!\Lambda\to pK^{+}p\pi$ process and detects all charged
particles in the final state, thus it is the most important subsystem
for this analysis. It consists of 2704 individual straw tubes combined
into 13 double layers normal to the beam axis with 3 azimuthal
orientations. The spatial resolution achieved under experimental
conditions of the individual straws has been shown to be
$\sigma\approx 150$\,\mum{} with an efficiency better than $\approx
98$\%. For the $p\Kaonp\Lambda$ final state this results in
a resolution of $\sigma_{m}\approx 1.1$\,\MeVcc{} for the $p\Lambda$
invariant mass at a reconstruction efficiency times acceptance of
$25$\%\cite{mythesis} for the charged decay mode.

\subsection{Data Analysis}
\label{sec:data-analysis}

\begin{figure}
  \centering
  \includegraphics[width=0.9\linewidth]{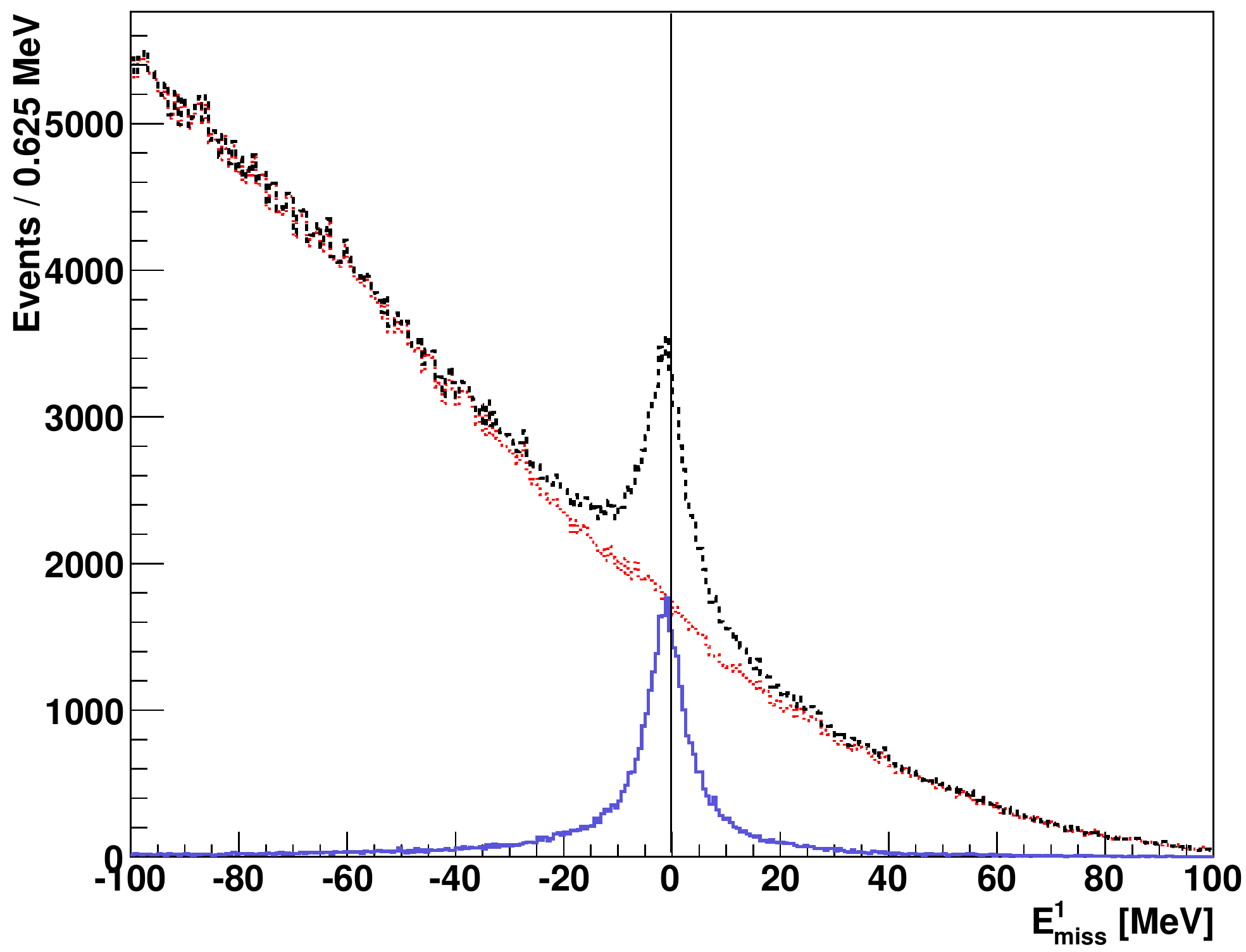}
  \caption{Missing energy spectrum at the primary vertex. From all
    events with a successful kinematic fit (dashed line) a sub-sample
    is selected (blue solid line) as described in the text. Also the
    complementary sub-sample is shown (red dotted line).}
  \label{fig:signalpeak}
\end{figure}

The primary signature of a good event is the combination of two
primary tracks from the target and two tracks from the delayed weak
decay of the $\Lambda$ particle. Additionally, the $\Lambda$ decay
plane contains the primary vertex position. After selecting events
that fulfill these criteria, a kinematic fit is performed which
minimizes the $\chisquare_{\text{kin}}$ with respect to the measured
track to wire distances in the STT. To reject background processes a
threshold is set on the reduced chi-square:
\begin{equation}
  \label{eq:1}
  \chisquare_{\text{kin}}/\text{NDF} < 5 \ .
\end{equation}
Additionally, events with a minimum distance $s_{\Lambda}$ between
the production and decay of the $\Lambda$ are selected
\begin{equation}
  \label{eq:2}
  s_{\Lambda} > 3\,\text{cm} \ .
\end{equation}
Furthermore, the laboratory angle between the $\Lambda$ and its decay
proton is required to fulfill the condition:
\begin{equation}
  \label{eq:3}
  \measuredangle(\Lambda, p) > 3\,\text{mrad} \ ,
\end{equation}
in order to reduce instrumental background from events with multiple
primary tracks.

The effectiveness of these selection criteria is evaluated by the
distribution of the missing energy ($E_{\text{miss}}^{1}$) at the
primary vertex with respect to a $\pkl{}$ final-state hypothesis,
before the kinematic fit. This is shown with the dashed line in
Fig.~\ref{fig:signalpeak} for all fit events with
$s_{\Lambda}>1\,\text{cm}$. The signal peak around $0\,\MeV$ lies on
top of a continuum from the instrumental background. By applying the
criteria in Eqs.~(\ref{eq:1}-\ref{eq:3}) only events in the peak are
selected (solid). At $p_{\text{beam}}=2.95\,\GeVc$ a total sample of
about $42,000$ events is obtained.

Physical background remains from the process\linebreak
\mbox{\pptopkstopklgamma}, where the unmeasured $\gamma$ carries away
$\approx 77\,\MeV{}$. This results in a deflection of the $\Lambda$ by
$\approx 2\degree{}$ in the laboratory frame. The event topologies are
therefore similar. Studies of Monte Carlo (MC) generated events have
shown that the contamination of the event sample under the
conditions in Eqs.~(\ref{eq:1}-\ref{eq:3}) is $\leq5\%$. Therefore, it is
neglected in the following analysis.

\subsection{Determination of the \boldmath{\Kaonp{}} Analyzing Power \boldmath{$A_{N}$}}
\label{sec:determination-a_n}

The analyzing power $A_{N}$ is a measure for the left/right asymmetry
$\epsilon_{LR}$ of the \Kaonp{} differential cross section and is
defined as:
\begin{equation}
  \label{eq:35}
  A_{N}(\theta^{*}_{\!K})\equiv \frac{\epsilon_{LR}(\theta^{*}_{\!K}, \phi)}{\cos(\phi) \cdot
  P} \ .
\end{equation}
with the beam polarization $P$. The asymmetry is determined from
\begin{equation}
  \label{eq:36}
  \begin{split}
    \epsilon_{LR}(\theta^{*}_{\!K}, \phi) = \frac{L(\theta^{*}_{\!K},
      \phi)-R(\theta^{*}_{\!K}, \phi)}{L(\theta^{*}_{\!K}, \phi)+R(\theta^{*}_{\!K},
      \phi)} \ , \\\phi\!\in\!\left(-\frac{\pi}{2},
  +\frac{\pi}{2}\right) \ ,
  \end{split}
\end{equation}
where
\begin{equation}
  \begin{split}
    L(\theta^{*}_{\!K}, \phi) &= \sqrt{N^{+}(\phi)\cdot N^{-}(\phi+\pi)}\\
    \text{and}\quad R(\theta^{*}_{\!K}, \phi)
    &= \sqrt{N^{-}(\phi)\cdot N^{+}(\phi+\pi)} \ .
    \label{eq:37}
  \end{split}
\end{equation}
Here, $N^{\pm}(\phi)$ is the number of events with spin up ($+$) and
spin down($-$) projectiles at the azimuthal angle $\phi$. The spin
direction was flipped after every extraction cycle (120\,s). By
multiplying the number of events on the opposite sides of the detector
and opposite spin states, systematic effects from asymmetries in the
detector acceptance are canceled to first order. The data has been
divided into eight bins in the \Kaonp{} azimuthal angle $\phi$.

The beam polarization was determined with the known analyzing power
and the measured asymmetry in \mbox{$\pol p p\to pp$} elastic
scattering. As a result we obtain $P = (61.0\pm 1.7)$\,\%. For that
the $pp$ analyzing power was taken from the SAID partial wave
analysis\cite{Arndt:2000xc}. The polar angular dependence is in good
agreement with SAID and with a previous measurement by
EDDA\cite{Altmeier:2000pe}. 

Possible systematic effects from different magnitudes of the + and -
beam polarization were investigated by measuring both quantities
independently. Within the experimental precision the two results,
$P_{+} = (66 \pm 4)\%$ and $P_{-} = (57\pm 4)\%$, are compatible. An
analysis of $A_{N}$ using $P_{\pm}$ separately for the corresponding data
samples yields a systematic deviation to the analyis with
Eq.~(\ref{eq:35}) of less than 30\% of the statistical precision.
Therefore, the difference is neglected in the following analysis.

\section{Results}
\label{sec:results}

\subsection{Dalitz Plot}
\label{sec:dalitz-plot}

\begin{figure}
  \centering \includegraphics[width=\linewidth]{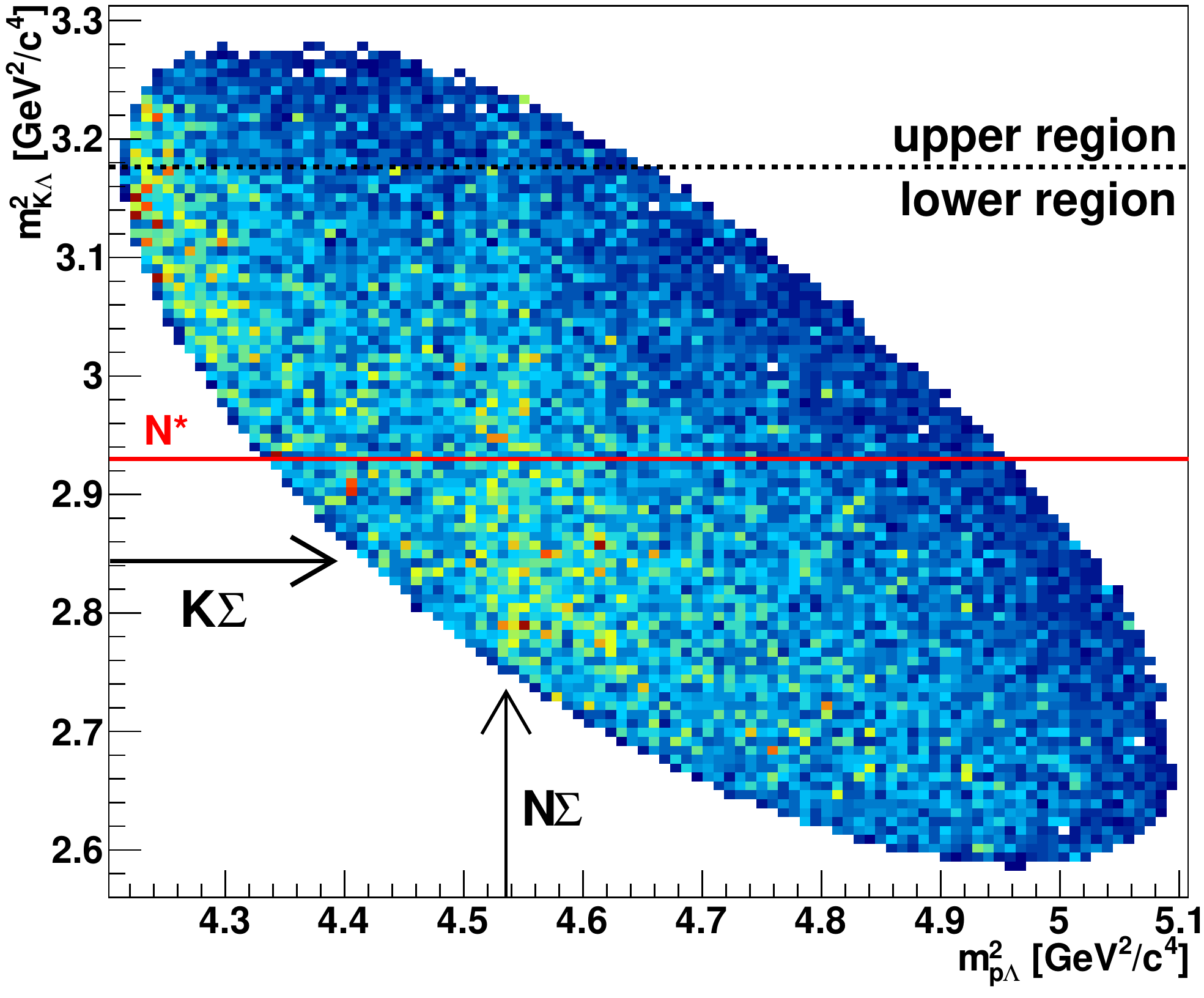} 
\caption{ The
 Dalitz plot of the reaction. Lighter colors indicate higher yield
 densities. The thresholds of the $N\!\Sigma$ and $K\!\Sigma$ channels are
 indicated by arrows, respectively. The region of the $N$(1710) and
 $N$(1720) resonances is indicated by a solid line. The dashed line
 marks the partition of the spectrum applied for the analysis
 discussed in Sec.~\ref{sec:effect-scatt-length} }  \label{fig:dalitz}
\end{figure}

The Dalitz plot of the selected event sample is shown in
Fig.~\ref{fig:dalitz}. It is corrected for the detector acceptance
with MC generated events. The complete kinematic acceptance of the
COSY-TOF detector is evident. The Dalitz plot density is strongly
enhanced at $m_{p\Lambda}^{2}=4.53\,\GeV^{2}/c^{4}$, i.e. the $N\kern
-0.15 em \Sigma$ threshold. This has been observed
before \cite{Siebert:1994jy, Tan:1969jq} and is usually interpreted as
an $N\kern -0.15 em \Sigma$--$p\Lambda$ coupled channel effect. The
high resolution available in this measurement makes an analysis of the
shape, position and strength of this structure interesting; however
that is beyond the scope of this report. It is analyzed in more detail
in Refs.~\cite{ElSamad:2012kg, Machner:2013hs}.

The enhancement of the production cross section close to threshold
from $p\Lambda$ interactions, as discussed in Sec.~\ref{sec:spintripl},
is clearly visible at low $m_{p\Lambda}$ values. The increasing
differential cross section for decreasing $m_{\kl}$ (see
Fig.~\ref{fig:dalitz}) can be explained by the influence of the
resonances $N\kern -0.1 em (1710)$ and/or $N\kern -0.1 em
(1720)$\cite{AbdElSamad:2010tz, AbdelBary:2010pc}.  In the Dalitz plot
these are located around $m_{\kl}^{2}\approx 2.93\,\GeV^{2}\!/c^{4}$.
However, due to their width of more than 100\,\MeVcc{} they do not
appear as narrow structures. For a theoretical description see, e.g.,
Refs.~\cite{oset:2011, gasp:2000}.

\subsection{Effective Scattering Length}
\label{sec:effect-scatt-length}

\begin{figure}[btp]
  \centering
  \includegraphics[width=\linewidth]{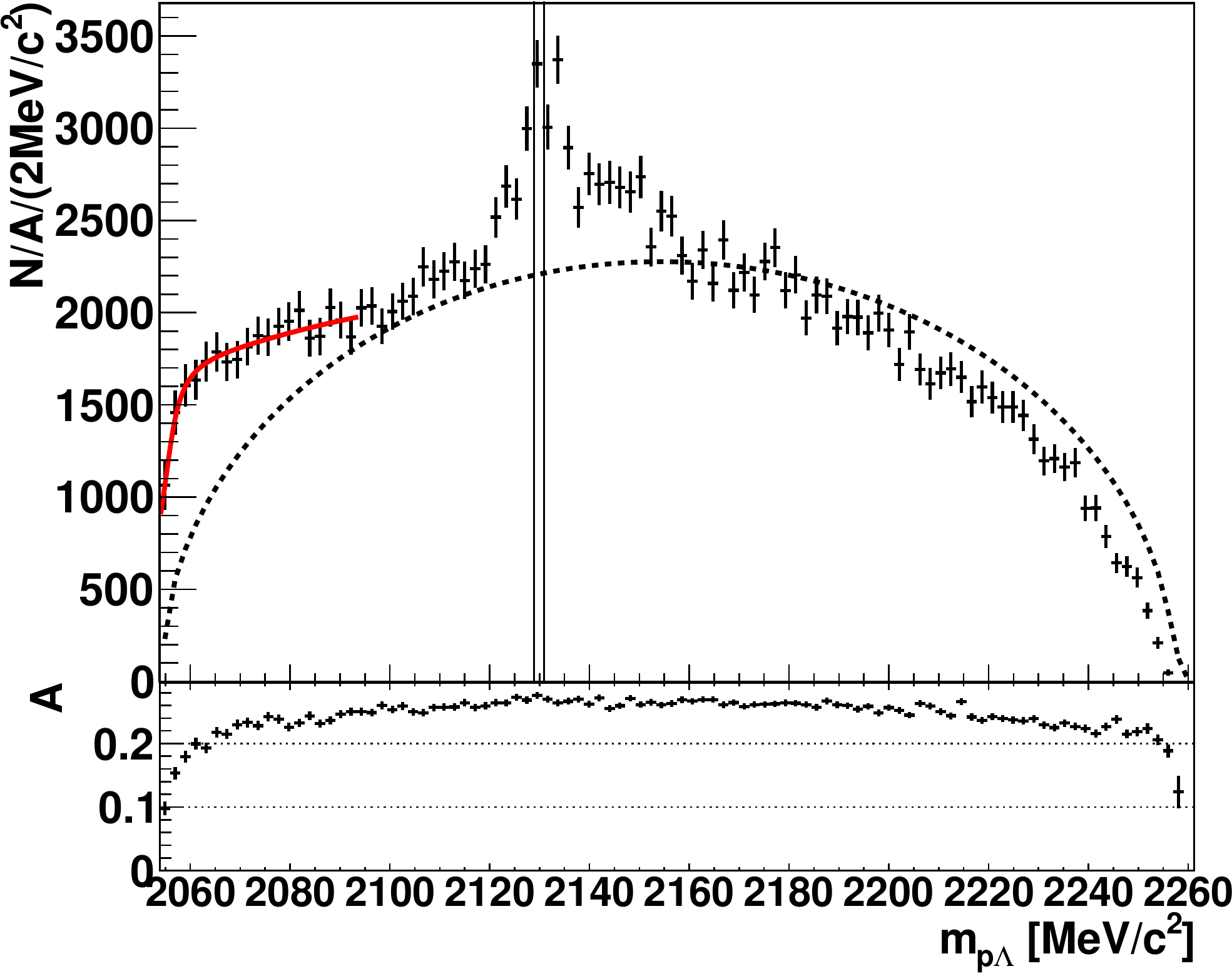}
  \caption{The spectrum of $m_{p\Lambda}$ corrected for acceptance (A)
    as it is given on the bottom. The two vertical lines indicate the
    $N\kern -0.15 em \Sigma$ thresholds. An arbitrarily scaled phase
    space distribution (dashed line) is shown to guide the
    eye. The solid line is a fit to the data as described in the
    text.}
\label{fig:mplspec}
\end{figure}

In Fig.~\ref{fig:mplspec} the $p\Lambda$ invariant mass spectrum is
shown. Since the time integrated luminosity of the event sample is not
needed for this analysis, only the number of measured events ($N$)
scaled with the detector acceptance times reconstruction efficiency
($A$) is given. The quantity $A$ has been determined with Monte Carlo
studies and is included at the bottom of the figure. It is noteworthy,
that the detector acceptance is nearly constant over a wide
$m_{p\Lambda}$ range but varies between 27\,\% and 10\% close to
threshold.

For comparison, an arbitrarily scaled three-body S-wave phase-space
distribution is shown with a solid line. The $N\kern -0.15 em \Sigma$
threshold positions of \mbox{$m_{n\Sigma^{+}}=2128.9\,\MeVcc$} and
\mbox{$m_{p\Sigma^{0}}=2130.9\,\MeVcc$} are marked with two vertical
lines. In this region a strong enhancement is visible.

There is also a sizable enhancement of the invariant mass spectrum at
$m_{p\Lambda}$ values close to the $p\Lambda$ threshold, i.e. in the
region relevant for the determination of the $p\Lambda$ scattering
length.  To test the method for the extraction of the scattering
length the measured invariant mass spectrum is fit with the function
given in Eq.~(\ref{eq:103}), convoluted with the detector resolution.
Within the range of $m_p+m_\Lambda \leq m_{p\Lambda} \leq
m_p+m_\Lambda+50$ $\MeVcc$ the achieved $\chisquare/\text{NDF}$ is
$0.32$. The corresponding result is indicated by the solid line in
Fig.~\ref{fig:mplspec}.

From a likelihood analysis\cite{Gasparyan:2003cc, mythesis} of the
highly correlated parameters we obtain \mbox{$a_{\rm
eff}=(-1.25\pm0.08\pm0.3)$\,fm}. Here, in the second and third term
the uncertainties from statistics and of the theoretical method are
given, respectively. Because the incoherent sum of the spin-singlet
and spin triplet $p\Lambda$ final-state interactions enter into the
production amplitude and their relative weights are unknown, $a_{\rm eff}$
is referred to as effective scattering length. Note, that it is not a
spin average.

A variation of the upper limit of the fit range between \mbox{(40,
50)\,\MeVcc{}} yields a stable result within the statistical error.
At an upper limit of 60\,\MeVcc{} the absolute value of the effective
scattering length increases by $\approx0.3$\,fm. Although this is
still within the systematic error, it might be connected to an
increasing importance of higher partial waves in the $\{p\Lambda\}$
system or it could signal already the onset of distortions caused by
the nearby $N\kern -0.15 em \Sigma$ threshold.

\begin{figure}
  \centering \includegraphics[width=\linewidth]{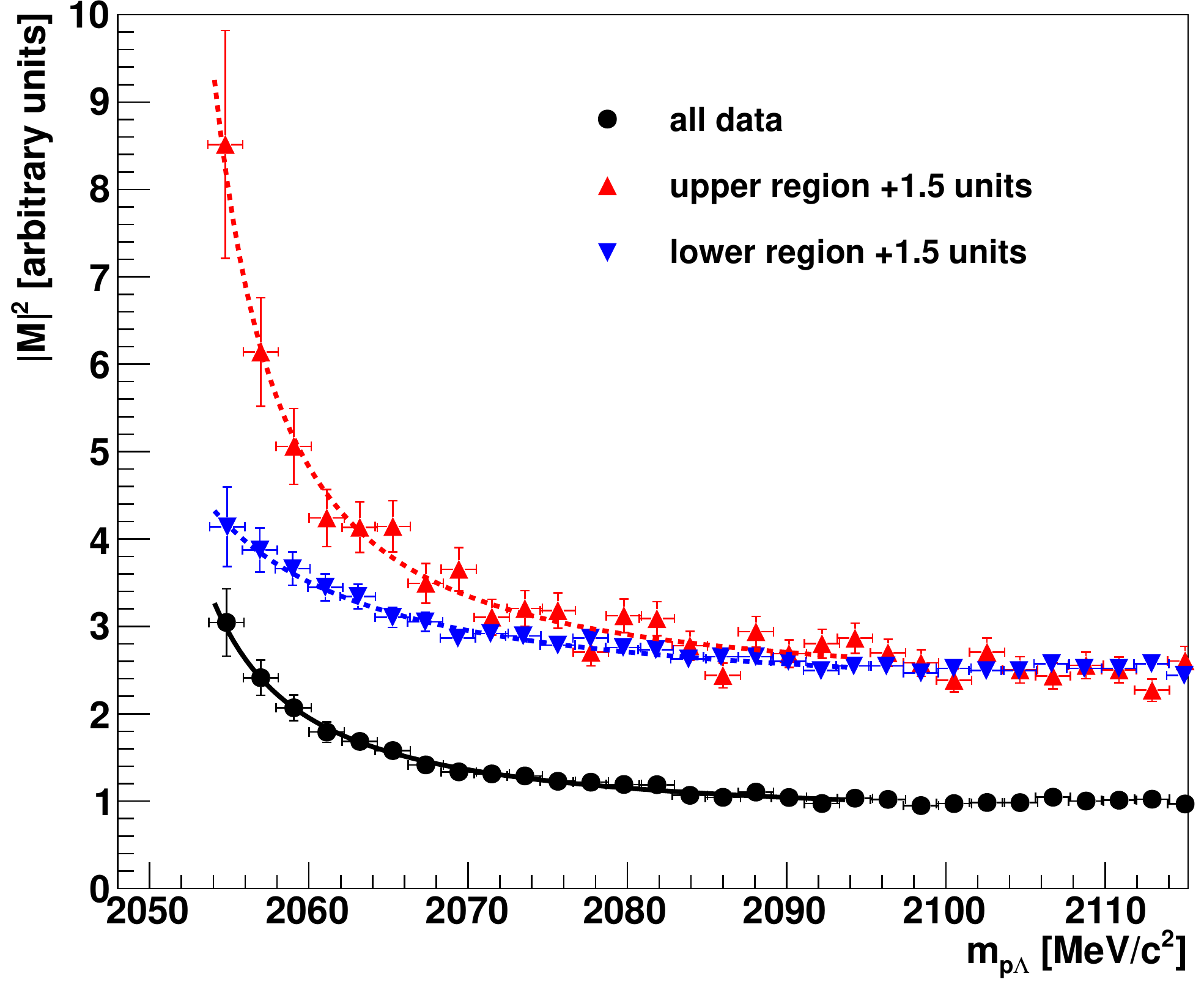}
   \caption{The quantity $|M|^{2}$ (circles) fit with the exponential
      function of Eq.~\eqref{eq:103} (solid line).  Results based on
      data from the upper (triangles up) and lower $m_{K\Lambda}$
      region (triangles down) of the Dalitz plot alone, as described
      in the text, are also shown.  For better readability the latter
      two are shifted +1.5 units on the y-axis,
      respectively.}  \label{fig:fsi}
\end{figure}

As emphasized in Ref.~\cite{Gasparyan:2003cc}, the error due to the
possible excitation of resonances or, more generally, of a final-state
interaction in the $K\Lambda$ and/or $KN$ subsystems cannot be
estimated on general grounds and, therefore, is not included in the
theoretical uncertainty of $0.3$ fm cited above. It can only be
quantified via a careful analysis of the Dalitz plot. 

For this purpose Fig.~\ref{fig:fsi} presents the ratio of the spectrum
to the arbitrarily scaled phase space (circles), see Eq.~(\ref{eq:7}).
Our fit based on the exponential function of Eq.~\eqref{eq:103} is
shown by the solid line. The data are normalized to have an average
$|M(m_{p\Lambda})|^{2}=1$ in the range
\mbox{$(2090<m_{p\Lambda}<2110)\,\MeVcc$}. To quantify the influence of the \nstar{} resonances on the measurement,
we apply the method described above to two separate
$m_{K\Lambda}$-regions of the Dalitz plot, namely \mbox{$(2.590 \leq
m_{\kl}^{2} \leq 3.176)\,\GeV{}^{2}\!/c^{4}$} (triangles down)
and \mbox{$(3.176 \leq m_{\kl}^{2} \leq 3.287)\,\GeV{}^{2}\!/c^{4}$}
(triangles up). These two regions are referred to as the lower ($\ell
r$) and upper ($ur$) ranges, respectively. Their boundaries are chosen
such that both regions include $m_{p\Lambda} = m_{p} + m_{\Lambda}$,
i.e. the near-threshold region relevant for the determination of the
$p\Lambda$ scattering length. To improve readability the data are
shifted by +1.5 units along the y-axis in Fig.~\ref{fig:fsi}.

It can be seen that the measured strength and shape of the final-state
interaction varies significantly between all three samples. We
obtain \mbox{$a_{\rm eff}^{\ell r}=(-0.86\pm0.06\pm0.3)$\,fm} for the
lower and
\mbox{$a_{\rm eff}^{ur}=(-2.06\pm0.16\pm0.3)$\,fm} for the upper range,
respectively. The difference of 1.20\,fm shows that the systematic
effect of \nstar{} resonances severely limits the precision of the
determination of $a_{\rm eff}$ from our data. This result agrees with
our observations for the Dalitz plot in Fig.~\ref{fig:dalitz}: At low
$m_{p\Lambda}^{2}$ the density clearly deviates from a homogeneous
distribution along $m_{K\Lambda}^{2}$, leading to a tilted shape of
the enhancement from final-state interactions.

\subsection{$\boldsymbol{\Kaonp{}}$ Analyzing Power}
\label{sec:spin-tripl-extr}

\begin{figure}
  \centering
  \includegraphics[width=\linewidth]{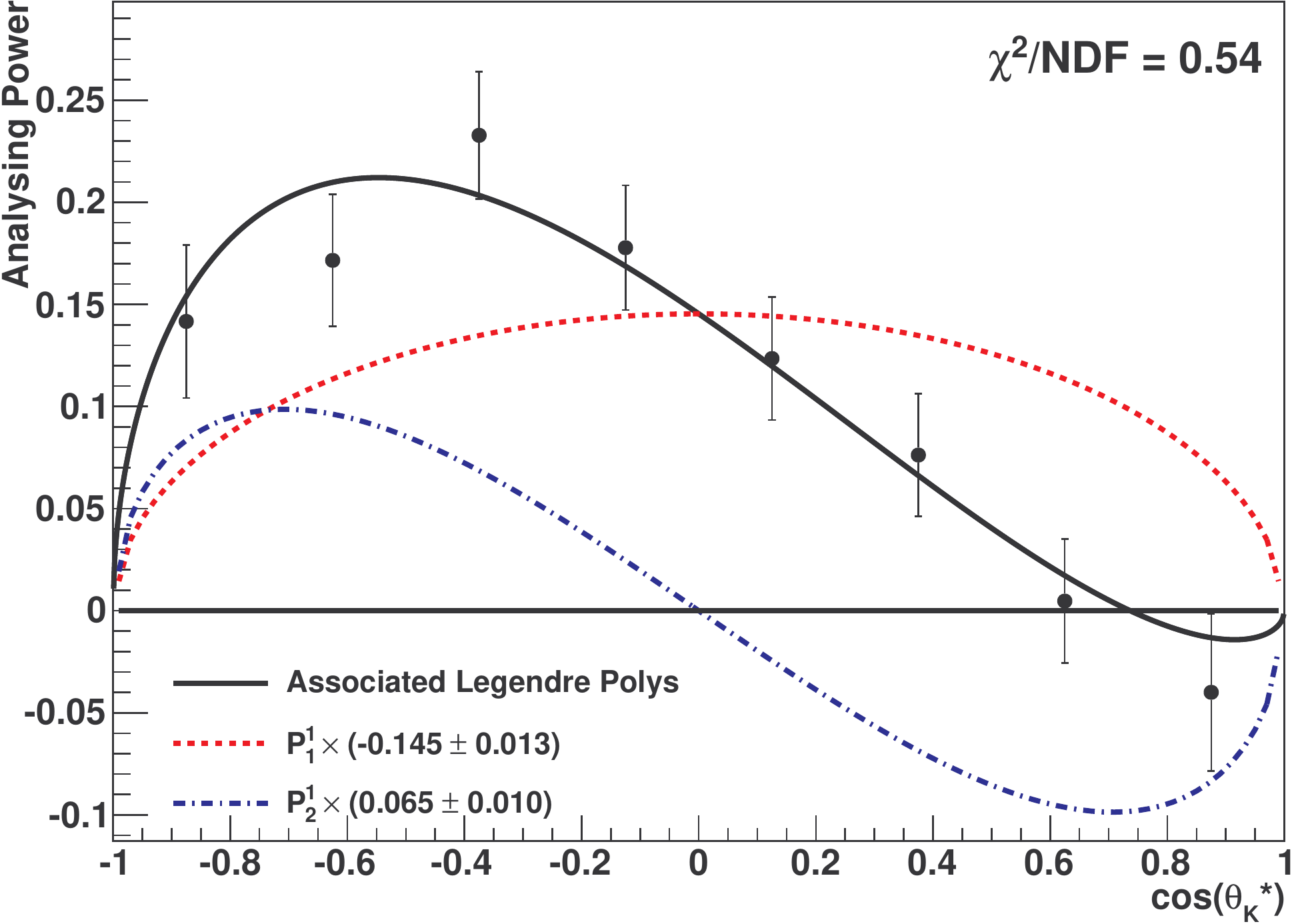}
  \caption{The \Kaonp{} analyzing power for the whole $m_{p\Lambda}$
    range. The fit (solid line) includes the associated Legendre
    polynomials ($P_{1}^{1}$) (dotted line) and ($P_{2}^{1}$)
  (dash-dotted line).}
  \label{fig:kaonanapow}
\end{figure}

In Fig.~\ref{fig:kaonanapow} the analyzing power of the $K^{+}$ is
shown for the whole $m_{p\Lambda}$ range. The parameters $\bar \alpha$
and $\bar \beta$ of Eq.~(\ref{eq:401}) are fit to the data (solid
line). The good quality of the fit, $\chisquare/\text{NDF} = 0.54$,
justifies the exclusion of higher order contributions in
Eqs.~(\ref{eq:101}) and (\ref{eq:401}). The symmetric (dotted line) and
antisymmetric contributions (dash-dotted line) are also shown
separately.

\begin{figure}
  \centering 
  \includegraphics[width=\linewidth]{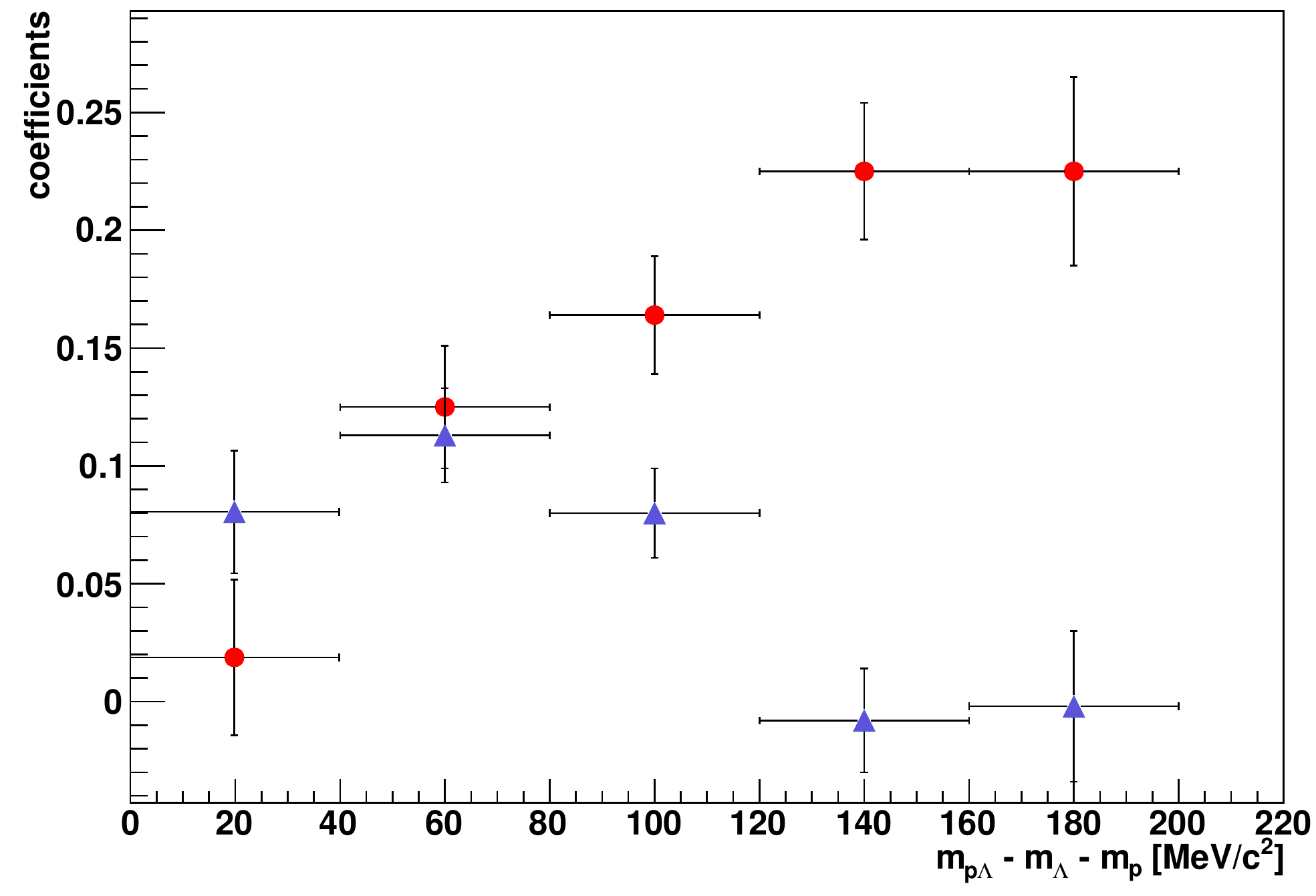} 
  \caption{The coefficients $\bar\alpha$ and $\bar\beta$ corresponding to the 
  forward-backward symmetric (circles) and antisymmetric (triangles) 
  contributions to the \Kaonp{} analyzing power as a function of 
  $m_{p\Lambda}$ according to Eq.~(\ref{eq:401}).}  \label{fig:edep}
\end{figure}

In Fig.~\ref{fig:edep} the event sample is binned in $m_{p\Lambda}$
and the fit results for $-\bar \alpha(m_{p\Lambda})$ (circles) and
$\bar \beta(m_{p\Lambda})$ (triangles) are given, respectively. The
antisymmetric part of $A_{N}$ approaches zero for higher $m_{p\Lambda}$.
This is reasonable because in this region the \Kaonp{} has the lowest
momentum which could be insufficient for s- and d-wave interference.
The symmetric part on the other hand is close to 25\% on the high end
of the spectrum and reduces to $\leq 11$\,\% with $3\sigma$ confidence
on the low end. This means that with the present quantity of data the
dependence of $\bar \alpha$ on $m_{p\Lambda}$ cannot be determined with
sufficient precision to extract the spin-triplet scattering length
with Eqs.~(\ref{eq:103}-\ref{eq:46}).

\section{Discussion}
\label{sec:discussion}

The observed behavior of the analyzing power is unexpected because in
the reaction $pp\to d\pi^{\!+}$, that has only spin 1 in the final
state and consequently the same selection rules, a value around 25\,\%
for the symmetric contribution has been
observed\cite{Mathie:1983,Hutcheon:1990pi,Korkmaz:1991vd}.

In principle, such small values could point to a complete absence of
spin-triplet production in the reaction $pp\to pK^+\Lambda$.  As
discussed in Sec.~\ref{sec:spintripl}, in that case contributions
with even \Kaonp{} orbital angular momentum are zero at low
$m_{p\Lambda}$.  It would then follow that $a_{\rm eff}$ as determined
in the last Section practically coincides with the $^{1}S_{0}$
scattering length $a_{s}$.  However, one has to keep in mind that such
small or vanishing contributions with even \Kaonp{} orbital angular
momentum are only a necessary but not a sufficient condition for the
absence of spin-triplet production, see Appendix B
of \cite{Gasparyan:2003cc}.  Thus, based on the present experiment,
the identification of our $a_{\rm eff}$ with $a_{s}$ is purely
hypothetical.

Realistic interaction potentials of the coupled $N \Lambda$--$N\Sigma$
systems \cite{Rijken:1998yy,Tominaga:2001,Haidenbauer:2005zh,Haidenbauer:2013}
which describe $p\Lambda \to p\Lambda$ elastic
scattering\cite{Alexander:1969cx} and also the binding energy of the
hypertriton \cite{Nogga:2001ef,Nogga:2013pwa} predict $(-1.4 \leq
a_{t} \leq -1.7)$\,fm and $(-2.5 \leq a_{s} \leq
-2.9)$\,fm. Significantly smaller values of those scattering lengths,
e.g. $a_{s} \approx -1.9$\, fm as suggested by an investigation
performed at leading order in chiral effective field
theory \cite{Polinder:2006zh} are no longer supported by the recent
extension of this study to next-to-leading
order \cite{Haidenbauer:2013}.  Since the $N\kern -0.1 em (1710)$ and
$N\kern -0.1 em (1720)$ resonances lie in the lower $m_{\kl}$ region
considered it is likely that their effect is weaker in the upper
region. Therefore, if we interpret the extracted $a^{ur}_{\rm eff}$ as
a lower boundary for the spin-singlet scattering length it would be
still compatible with the theoretical picture, especially within the
uncertainties.  However, the value of $a^{ur}_{\rm eff}$ is more or
less halfway between the ranges predicted for the singlet- and triplet
$p\Lambda$ scattering lengths. This is also consistent with the
naive expectation for this quantity if both spin states are produced.

Our $a^{ur}_{\rm eff}$ is in good agreement with the value published
by the HIRES collaboration\cite{Budzanowski:2010ib} which is
\mbox{$a=-2.4^{+0.16}_{-0.25}$}. 
Their result is based on a combined analysis of FSI effects in the
reaction $pp \to pK^+\Lambda$ at $p_{\text{beam}}=2.7\,\GeVc$ and
$p\Lambda$ elastic scattering data. If one takes into account the
arguments of Ref.~\cite{Gasparyan:2005fk}, where it is demonstrated
that the method applied in Ref.~\cite{Budzanowski:2010ib}
overestimates the scattering length by approximately $0.4$\,fm, the
agreement is even better.  Certainly, the effect of \nstar{}
resonances on the HIRES result is unclear, even though earlier
measurements of
\mbox{COSY-TOF} indicate a reduced influence on the Dalitz
plot\cite{AbdElSamad:2010tz} at that beam momentum.  Note that the
authors of Ref.~\cite{Budzanowski:2010ib} argue that, based on their
analysis, the production of the $p\Lambda$ system in a spin-triplet
state is negligible.

As already said above, the behavior of the analyzing power as found in
our experiment does not rule out the presence of spin-triplet
$p\Lambda$ states: Since only the imaginary parts of the amplitudes
enter the analyzing power, an accidental phase cancellation is
possible. Definite conclusions can only be drawn if one can set
quantitative upper limits for the symmetric part of the analyzing
power. It is therefore important to collect higher statistics and
measure at different beam energies. It should also be investigated how
other polarization observables, e.g. the $\Lambda$ polarization, can
be employed to put quantitative constraints on the production of
$p\Lambda$ in a spin-triplet state.

\section{Conclusion}
\label{sec:conclusion}

The effective $p\Lambda$ scattering length has been determined from
final-state interactions in the reaction $\pptopkl$ at
$p_{\text{beam}}=2.95\,\GeVc$.  An examination of the influence of the
excitation of \nstar{} resonances in the $K\!\Lambda$ channel revealed
that they introduce a large uncertainty on the analysis. This should
be and has to be taken into account in any attempt to determine the
$p\Lambda$ scattering length from this reaction.  Whether the
influence of those resonances is only particularly strong at beam
momenta like those of the present experiment remains to be seen.  For
further studies, data from different beam momenta is highly
desirable. Eventually, this could allow to quantify and even control
the effect from resonances. It might even be possible to identify a
range of beam momenta where the presented method is not systematically
distorted.

The \Kaonp{} analyzing power has been measured as a function of the
$p\Lambda$ invariant mass. The vanishing symmetric contribution to the
analyzing power at low values of $m_{p\Lambda}$ prohibits the
extraction of the spin-triplet scattering length with the present
quantity of data. The hypothesis of exclusive spin-singlet production
of the $p\Lambda$ system was discussed as an explanation for this
unexpected behavior. For a decisive study, measurements with higher
statistics are needed. Especially, the possibilities to exploit the
$\Lambda$ polarization should be investigated.

\section{Acknowledgments}
\label{sec:acknowledgement}

The research leading to these results has received funding from the
European Union Seventh Framework Programme (FP7/2007-2013) under grant
agreement no. 283286. 

This work comprises part of the PhD thesis of Matthias Röder.

We would like to thank the COSY operation crew for providing excellent
quality beams.

\end{document}